\newif
\ifconfver
\confvertrue        
\ifconfver
    \documentclass[10pt,journal,final,twoside]{IEEEtran}
\else
    \documentclass[11pt,draftcls,onecolumn]{IEEEtran}
\fi
\usepackage{placeins,float,bbm,xspace,empheq,fancybox,amsmath,amssymb,graphicx,epstopdf,epsfig,subfigure,syntonly,times,amsthm} 
\usepackage{psfrag,color,bm,array,cite}
\usepackage{url,cite,footnote,xspace,syntonly,algorithm,algorithmic}
\usepackage{verbatim,multirow,slashbox}
\usepackage[T1]{fontenc}
\usepackage{algorithm,cite}

\allowdisplaybreaks[3]

\newtheorem{remark}{\bfseries Remark}
\input{mysymbol.sty}

\newcommand{\ml}{\color{black}{}}

\begin{document}


\title{A Dynamic Response Recovery Framework \\ Using Ambient Synchrophasor Data}
\author{
\IEEEauthorblockN{Shaohui Liu},~\IEEEmembership{Student Member, IEEE}, \IEEEauthorblockN{Hao Zhu},~\IEEEmembership{Senior Member, IEEE}, and \IEEEauthorblockN{Vassilis Kekatos},~\IEEEmembership{Senior Member, IEEE}

\thanks{\protect\rule{0pt}{3mm} 
	This work has been supported by NSF Awards ECCS-1802319 and 1751085. 

	S. Liu, and H. Zhu are with the Department of Electrical \& Computer Engineering, The University of Texas at Austin, 2501 Speedway, Austin, TX 78712, USA; e-mail: {\{shaohui.liu, haozhu\}{@}utexas.edu}.

	V. Kekatos is with the Bradley Dept. of ECE, Virginia Tech, Blacksburg, VA 24061, USA; e-mail: {kekatos{@}vt.edu}.
}}

\markboth{(REVISED)}%
{Liu \MakeLowercase{\textit{et al.}}: A Dynamic Response Recovery Framework Using Ambient Synchrophasor Data}
\renewcommand{\thepage}{}
\maketitle
\pagenumbering{arabic}

%
\begin{abstract}
Wide-area dynamic studies are of paramount importance to ensure the stability and reliability of power grids. Recent deployments of synchrophasor technology have rendered data-driven modeling possible by analyzing fast-rate dynamic measurements. This paper puts forth a comprehensive framework for inferring the dynamic responses of the power system in the small-signal regime using ubiquitous ambient data collected during normal grid operations. We have shown that the impulse response between any pair of locations can be recovered within a scaling ambiguity in a model-free fashion by simply cross-correlating synchrophasor data streams collected only at those two locations. The result has been established via model-based analysis of linearized second-order swing dynamics under certain conditions. Nevertheless, numerical tests demonstrate that this novel data-driven approach is applicable to realistic power system models including nonlinear higher-order dynamics and controller effects. Its practical value is corroborated by excellent recovery performance attained in the WSCC 9-bus system and a synthetic 2000-bus Texas system.
\end{abstract}




\section{Introduction}\label{sec:intro}

Power system dynamic studies are critical for maintaining system stability and achieving secure decision makings at control centers\cite[Ch.~1]{kundur1994power}. During correlated failures of multiple components, power imbalance can quickly propagate throughout an interconnection as result of the so-termed electro-mechanical (EM) oscillations \cite{thorp1998electromechanical,backhaus2012electromechanical}. 
If the oscillation modes are poorly damped, a small input disturbance could trigger loss of synchronization and even cascading outages at other areas, such as the 1996 US/Canada Western Interconnection blackout and the 2002 Italian blackout; see e.g., \cite{blackout2005}. Thus, enhancing the modeling of dynamic responses is of imperative needs.

In recent years, we have witnessed the rise of synchrophasor technology and wide deployment of phasor measurement units (PMUs). The high-rate synchrophasor data samples on bus frequency/angle and line flow provide unprecedented visibility regarding the transient behavior of power systems. Using synchrophasor measurements, a data-driven framework has been advocated for estimating the model parameters of dynamical components such as generators or loads \cite{huang2013generator,zhang2016dependency}, or for directly constructing the network dynamic equations \cite{chavan2016identification}. Most of these data-driven approaches rely on significant transient responses to large system faults. Thus, they are limited to post-event analysis and cannot incorporate the more ubiquitous type of   ambient synchrophasor data. 

Ambient data are highly relevant for the \textit{small-signal analysis} in power system dynamics \cite[Ch.~12]{kundur1994power}. By linearizing the nonlinear dynamic model around an operating point, one can characterize the stability margin and determine the inertia or primary frequency responses \footnote{{\ml In this paper, frequency, angel, or line flow responses refer to the impulse responses of frequency, angel, or line flow in the time domain.}}. 
Hence, ambient synchrophasor data have been popularly used for estimating the oscillation modes; see e.g., \cite{pierre1997initial,zhou2008electromechanical,zhou2009electromechanical,ning2014two,wu2016fast} and references therein. To recover system dynamic responses, the statistical information of ambient data has been utilized to estimate the dynamic state Jacobian matrix \cite{wang2017pmu,ramakrishna2021grid,sheng2020online}. However, these approaches require the availability of state measurements at a majority of bus locations and  cannot cope with limited PMU deployment thus far. In addition to dynamic modeling, ambient data have also been used for quantifying voltage stability metrics \cite{chevalier2018mitigating}.

The goal of the present work is to recover the inertia-based dynamic system responses in the small-signal regime from ambient synchrophasor measurements. We propose a cross-correlation based approach to process ambient synchrophasor data. The proposed cross-correlation approach is very general and flexible in the data types or PMU locations, as it can incorporate any frequency, angle, or line flow data streams from two arbitrary locations. Targeting at the small-signal analysis, we first study the well-known second-order dynamics for this regime to establish the theoretical equivalence for the proposed data-driven approach. This equivalence needs homogeneous damping among significant inter-area modes, which is reasonable for a wide-area system \cite{cui2017inter} and as corroborated by numerical tests later on. Accordingly, an ambient data-driven framework is developed to recover the frequency, angle, or line flow responses to a disturbance from any input location. Going beyond the theoretical equivalence, the effectiveness of the proposed framework is numerically demonstrated by realistic power system models that include higher-order dynamics and controller effects. 

The present work significantly extends our earlier work \cite{huynh2018data} from ambient frequency data analytics to a comprehensive framework encompassing ambient data of angles and line flows. The latter is typically of higher accuracy than frequency, {\ml as real-world PMUs produce frequency data using low-pass filters to eliminate certain dynamics \cite{pmu_report2020}. Because ambient signals have very small variations around the nominal values, these filtering processes can significantly affect the quality of frequency data, speaking for the importance of extending to angle and line flow data.}  Another related work \cite{jalali2021inferring} has recently proposed a Gaussian Processes (GP) based approach for inferring data streams at multiple locations, at a slightly higher computation complexity. Thanks to the data-driven nature, the proposed methods can be conveniently applied to {\ml obtain the frequency representation of oscillation modes, which are very useful for evaluating the effectiveness of advanced control designs especially in the presence of new grid components like inverter-based resources.} To sum up, the main contribution of this work is two-fold: 
\begin{enumerate}
    \item Establish the equivalence between cross-correlation of various ambient data and the corresponding  angle or line flow responses to an arbitrary disturbance input. This equivalence builds upon the small-signal second-order dynamics and considers reasonable assumptions for wide-area systems; and 
    \item Develop a fully data-driven framework to recover the system responses that can incorporate all types of data at minimal PMU deployment. The proposed recovery algorithm requires no knowledge of the actual system model or parameters and is applicable to synchrophasor data streams from any pair of two locations. 
    {\ml \item Validate the proposed methods under a more realistic ambient condition by perturbing all load demands than the approach in \cite{huynh2018data} of perturbing generator inputs. We have also tested on synthetic PMU data which include the filtering effect and measurement noises as in actual meters, and the results further corroborate the practical values of using ambient angle and line flow data. }
\end{enumerate}

The rest of paper is organized as follows.  Section \ref{sec:ps} formulates the problem by defining ambient conditions. Section \ref{sec:model} establishes the equivalence results between small-signal dynamic responses and the cross-correlation of ambient data. Accordingly, Section \ref{sec:algorithm} develops the proposed data-driven algorithm based on cross-correlation and Section \ref{sec:numerical_results} demonstrates its validity on realistic dynamic models including the WSCC 9-bus system and a carefully validated 2000-bus Texas system. Section \ref{sec:con} concludes this paper.


\section{Problem Statement}
\label{sec:ps}


\begin{table}[t!]
\centering
\begin{tabular}{l l l} 
 \hline
 Notation & Description & Type \\ [0.5ex] 
 \hline\hline
 $\hat{\delta}_\ell,\hat{\omega}_\ell$ & generator $\ell$'s rotor angle and speed   & ambient data \\
 $\hat{\theta}_i,f_{ij}$ & phase angle of bus $i$, flow of line $(i,j)$ & ambient data \\
 $\delta_\ell,\omega_\ell$ & generator $\ell$'s rotor angle and speed & system variable \\
 $\theta_i, p_{ij}$ & phase angle of bus $i$,  flow of line $(i,j)$ & system variable \\
 $u_\ell$ & generator $\ell$'s active power schedule & system variable\\
 $\bbM,\; \bbD$ & generator inertia, damping coefficients& system parameter \\
 $\bbK$ & power flow Jacobian matrix & system parameter \\
 [1ex] 
 \hline \\
\end{tabular}
\caption{List of Symbols}
\label{table:1}
\vspace*{-5mm}
\end{table}

The dynamics of a power system can be generally described by a set of nonlinear differential and algebraic equations (DAEs), as given by \cite[Ch. 6-9]{arthur2000power}
\begin{align}
\begin{cases}
    \Dot{\bbx} &= f(\bbx,\bby,\bbu) \\
    \mathbf 0 &= g(\bbx,\bby)
\end{cases}
\label{eq:dae}
\end{align}
where vector $\bbx$ includes all state variables such as rotor angle $\delta_\ell$, speed $\omega_\ell$, and exciter/governor status per generator $\ell$; the output vector $\bby$ consists of all algebraic variables such as voltage magnitude $V_n$ and phase angle $\theta_n$ per bus $n$; and the input vector $\bbu$ represents the dispatch signals for the active power schedule at all generators. 

Our goal is to infer the grid's dynamic responses under \textit{small-signal} disturbances from ambient synchrophasor data. Dynamic model \eqref{eq:dae} can be approximated by a linear time-invariant (LTI) system with all variables represented by the \textit{deviations} from their steady-state values. For simplicity, the term deviations will be dropped henceforth. Under this LTI approximation, the dynamic response is fully characterized by an input through the \textit{impulse response}. Let $T_{u_k,\delta_\ell}(\tau)$ denote the impulse response of the target $\delta_\ell$ from input source at $u_k$, and similarly for other target variables such as  $T_{u_k,\omega_\ell}(\tau)$ and $T_{u_k,\theta_n}(\tau)$.  
Note that the small-signal analysis could use simplified modeling considerations for model-based analysis as explored later on. 


We propose recovering the dynamic responses by cross-correlating synchrophasor data collected under ambient conditions. Such methodology does not need to know the system model or probe the system with any particular inputs. To define the ambient conditions, random perturbations of active power injection (due to load or generation variations) lead to a ``white-noise'' input $\bbu(t) = \bbnu(t)$ satisfying \cite{ning2014two,wang2017pmu,wang2015data}: 
\begin{align}\
	\mathbb{E}\left[\bbnu(t)\right] &= \mathbf 0 \notag\\
	\mathbb{E}\left[\bbnu(t)  \bbnu^\top(t-\tau) \right] &= \bbSigma \Delta(\tau)
	\label{eq:input_var}
\end{align}
where $\Delta(t)$ is the Dirac delta function. Under the input in \eqref{eq:input_var}, the corresponding ambient state/output will be denoted by the hat symbol, such as  $\hat{\delta}_\ell(t)$ and $\hat{\omega}_\ell(t)$. The cross-correlation of ambient angle signals is given by
\begin{align}
	C_{\hat{\delta}_k\hat{\delta}_\ell}(\tau) &\triangleq \lim_{T\rightarrow\infty} \frac{1}{2T} \int_{-T}^{T} \hat{\delta}_k(t)\hat{\delta}_\ell(t-\tau)dt \notag \\
	 &= \mathbb{E}\left[\hat{\delta}_k(t)\hat{\delta}_\ell(t-\tau)\right]
	\label{eq:cross_correlation}
\end{align}
where the second equality is due to the stationary input process [cf. \eqref{eq:input_var}], which makes the sample average  in the definition asymptotically equivalent to expectation  \cite[Ch.~9]{stirzaker1992probability}.
The same equivalence holds for other cross-correlation results, which will be used to recover the system dynamic responses. 

\begin{remark}{\emph{(States versus measurements)}}  \label{rmk:state}
	PMUs are installed at buses or branches to measure the electrical quantities of the grid, and do not directly observe generator states $\hat{\delta}_\ell$ or $\hat{\omega}_\ell$. Nevertheless, these state variables are highly related to their local grid-level measurements. Generator speed $\hat{\omega}_\ell(t)$ is well represented by the electric frequency at the generator bus $n$, namely $\frac{d\hat{\theta}_n(t)}{dt}$ \cite{markham2014electromechanical}. Hence, the grid-level frequency data is an excellent surrogate for speed states. Furthermore, the ambient rotor angle-based results can eventually be generalized to grid-level measurements such as bus phase angle or line flow, by utilizing the linearized relations as in small-signal analysis.  
\end{remark}

\section{Model-Based Analysis}
\label{sec:model}

We consider the small-signal analysis of dynamic responses using the classical second-order generator model \cite[Ch. 9]{arthur2000power}. The state only includes the rotor angle and speed (frequency) vectors, $\bbdelta, \bbomega \in \mathbb R^N$ for a total of $N$ generators. The second-order model could also be viewed as a simplified, reduced model representation under higher-order dynamics due to governor and excitation controls \cite[Ch.~12]{arthur2000power}. The linearized model follows the swing equation as
\begin{align}
\begin{cases}
    \Dot{\bbdelta} &= \bbomega \\
    \bbM\Dot{\bbomega} &= -\bbK\bbdelta - \bbD\bbomega + \bbu
\end{cases}
\label{eq:swing1}
\end{align}
where 
the diagonal matrices $\bbM$ and $\bbD$ contain respectively the generator inertia and damping constants, while  $\bbK$ is the power flow Jacobian matrix evaluated at the given operating point. 
The swing dynamics of \eqref{eq:swing1} can also be written in the equivalent second-order form: 
\begin{align}
	\bbM\Ddot{\bbdelta} + \bbD\Dot{\bbdelta} + \bbK\bbdelta = \bbu.
	\label{eq:swing2}
\end{align}

In order to analyze the theoretical properties of \eqref{eq:swing2}, the following assumptions are made to simplify and decouple the system into independent modes: 
\begin{assumption}
The generator inertia and damping constants are homogeneous; namely $\bbD = \gamma \bbM$ for a constant $\gamma >0$.
\label{assump1}
\end{assumption}
\begin{assumption}
The power flow Jacobian matrix $\bbK$ is a symmetric Laplacian matrix and positive semidefinite (PSD).
\label{assump2}
\end{assumption}
The condition in (AS\ref{assump1}) can hold if parameters of each generator are designed to scale proportionally to its power rating, which has been frequently adopted for approximating power system dynamics~\cite{Low18,Paganini19,huynh2018data}. If transmission lines are all purely inductive (lossless) and loads are of constant power outputs for frequency-only dynamics in \eqref{eq:swing1}, matrix $\bbK$ becomes exactly symmetric and (AS\ref{assump2}) would hold. 
{\ml Certain load models (e.g., constant-current ones \cite{osti_1004165}) could affect the dynamic models, but may have minimal impact on forming matrix $\bbK$. This is because under small-signal analysis, the system voltages tend to be steady, thus leading to minimal load power changes.}
Both assumptions are used for establishing the analytical results only and will be waived during the numerical tests; see Remark \ref{rmk:model} for further discussions on generalizability.

Under (AS\ref{assump1})-(AS\ref{assump2}), one can decouple the system in \eqref{eq:swing2} through a linear transformation $\bbdelta = \bbV \bbz$, where matrix $\bbV =[V_{ki}]_{1\leq k,i\leq N}$ is specified by the generalized eigenvalue problem $\bbK\bbV = \bbM\bbV\bbLambda$ and the diagonal matrix $\bbLambda$ has the $N$ eigenvalues $\lambda_{i} \geq 0$. In addition, $\bbV$ is $\bbM$-orthonormal and satisfies \cite[Sec. 5.2]{strang2006linear}:
\begin{align}
\bbV^\top \bbM\bbV = \bbI~~\textrm{and}~~\bbV^\top \bbK\bbV = \bbLambda. \label{eq:MK}
\end{align}
Substituting \eqref{eq:MK} into \eqref{eq:swing2} and utilizing (AS\ref{assump1}) lead to a completely decoupled second-order system, given by 
\begin{align}
    \Ddot{\bbz} + \gamma\Dot{\bbz} + \bbLambda \bbz = \bbV^\top \bbu. 
    \label{eq:swing_modes}
\end{align}
Solving for each independent mode $z_i$ in \eqref{eq:swing_modes} gives rise to the impulse responses as
\begin{align}
	T_{u_k,\omega_\ell}(\tau) &= \sum_{i=1}^N V_{ki}V_{\ell i} ~\eta_i \left(c_i e^{c_i\tau} -d_i e^{d_i\tau} \right) \label{eq:freq_impz}\\
	T_{u_k,\delta_\ell}(\tau) &= \sum_{i=1}^N V_{ki}V_{\ell i}~\eta_i \left( e^{c_i\tau} - e^{d_i\tau} \right)
	\label{eq:ang_impz}
\end{align}
%
with the mode-associated complex parameters: 
\begin{align*}
c_i &= \frac{-\gamma + \sqrt{\gamma^2 - 4\lambda_i}}{2},~~~
    d_i = \frac{-\gamma - \sqrt{\gamma^2 - 4\lambda_i}}{2},
\end{align*}
and the coefficient $\eta_i 
=\frac{1}{\sqrt{\gamma^2 - 4\lambda_i}}$ for $1\leq i \leq N$.
Note that since $\bbK$ is the Laplacian, it has one eigenvalue $\lambda_1 = 0$ which gives rise to a marginally stable mode $c_1=0$. This is known as the reference angle issue which will be eliminated by filtering out low-frequency components of the ambient data; see [S2] of Sec. \ref{sec:algorithm}. Without loss of generalizability (Wlog), we assume the system has been transformed to eliminate the zero eigenvalue and thus all modes are stable; see e.g., \cite[Ch. 12]{sauer_book} for more details on this step.   Upon obtaining the impulse responses in \eqref{eq:freq_impz}-\eqref{eq:ang_impz}, we will exploit the structure therein for ambient data processing.



\section{Ambient Data Analytics}
\label{sec:theoretical_results}

To compare the cross-correlation with model-based dynamic responses, we formally define the ambient conditions here.
\begin{assumption}
The ambient data during nominal operations is generated by random noise $\bbnu(t)$ that satisfies \eqref{eq:input_var} with variance proportional to inertia; i.e., $\bbSigma = \alpha \bbM$ with $\alpha>0$.
\label{assump3}
\end{assumption}
{\ml 
The assumption is introduced to guarantee that all modes in \eqref{eq:swing_modes} are equally and independently excited, thanks to the diagonalization $\bbV^\top\bbSigma \bbV = \alpha \bbI$ [cf. \eqref{eq:MK}]. Note that real-world power systems may not perfectly balance the generation inertia with load variability, as most types of generation are placed based upon resource availability. However, for a large interconnection (AS3) could hold broadly over all control areas, instead of at every location. Furthermore, to deal with lowering inertia in current power systems, the placement of virtual synchronous generators \cite{arghir2018grid} and virtual inertia \cite{poolla2017optimal,poolla2019placement} tends to account for load variability. To sum up, even though (AS3) may not hold perfectly for actual grids, it aims to ensure homogeneous damping among significant inter-area modes, which is reasonable for a wide-area system as shown by actual synchrophasor data analysis in \cite{cui2017inter}.
%
%
}. 

For ambient frequency data generated under (AS\ref{assump3}), \cite{huynh2018data} has established the following equivalence result for inferring frequency response.

\begin{lemma}{\emph{(Frequency Response)}}\label{prop:freq}
Under (AS\ref{assump1})-(AS\ref{assump3}), the cross-correlation of ambient frequency $\hat{\omega}_k$ and $\hat{\omega}_\ell$ is related to the frequency response as
\begin{align}
    T_{u_k,\omega_\ell}(\tau) = -\frac{2\gamma}{\alpha} C_{\hat{\omega}_k,\hat{\omega}_\ell}(\tau). \label{eq:freq} 
\end{align}
\end{lemma}

{According to Lemma \ref{prop:freq}, the cross-correlation between the ambient frequencies across two generator buses is proportional to the impulse response between those two buses, thus gives a model-free tool for processing frequency data to infer the grid impulse responses in frequency. In addition to frequency, the PMUs measure diverse types of quantities such as bus voltage angle and line active/reactive power. Hence, it will be very useful if this framework of processing ambient frequency data can be extended to general types of PMU measurements. Of course, one can always filter ambient angle data through differentiation to obtain the corresponding frequency data and use the latter for cross-correlation. Nonetheless, the accuracy of such a two-step process can be greatly affected by the filtering design and suffer significantly from measurement noise in the small-signal regime. Instead, we propose several approaches for directly processing ambient angle and power measurements as provided by PMUs.} 


\begin{proposition}{\emph{(Angle Response)}}\label{prop:ang}
	Under (AS\ref{assump1})-(AS\ref{assump3}), the cross-correlation of ambient angle $\hat{\delta}_k$ and $\hat{\delta}_\ell$ is related to the angle response as
\begin{align}
T_{u_k,\delta_\ell}(\tau) &= -\frac{2\gamma}{\alpha} \frac{d}{d\tau}C_{\hat{\delta}_k,\hat{\delta}_\ell}(\tau)= -\frac{2\gamma}{\alpha} C_{\hat \omega_k,\hat\delta_\ell}(\tau) \label{eq:ang_resp} 
\end{align}
\end{proposition}
\begin{IEEEproof}
The ambient angle is the convolution of input noise $\bbnu(t)$ and the impulse response  in \eqref{eq:ang_impz}. Hence, we can define the vector $\bbh_k(t) = \left[ V_{ki}\eta_i \left( e^{c_it} - e^{d_it} \right) \right]_{N\times1}$ and show that 
\begin{align*}
    &C_{\hat{\delta}_k,\hat{\delta}_\ell}(\tau) 
    = \int_0^\infty dt_1 \int_\tau^\infty dt_2 \nonumber\\
    & \;\;\;\;\;\;\;\;\;\;\;\;\;\;\;\bbh_k(t_1)^\top \bbV^\top \mathbb{E}\left[\bbnu(t-t_1)\bbnu(t-\tau-t_2)^\top\right] \bbV \bbh_\ell(t_2) \nonumber\\
    =& -\alpha\sum_{i=1}^N V_{ki}V_{\ell i}\eta_i^2 \;\left[ \left(\frac{1}{2c_i}+\frac{1}{\gamma}\right)e^{c_i  \tau} 
    + \left(\frac{1}{2d_i}+\frac{1}{\gamma}\right)e^{d_i  \tau} \right]
\end{align*}
where the second equality uses the white-noise property  and the diagonalization $\bbV^\top\bbSigma \bbV = \alpha \bbI$ in (AS\ref{assump3}). Taking its derivative and utilizing the relations among $c_i$, $d_i$ and $\eta_i$ lead to the equivalence between $\frac{d}{d\tau}C_{\hat{\delta}_k,\hat{\delta}_\ell}(\tau)$ and $T_{u_k,\delta_\ell}(\tau)$ as in \eqref{eq:ang_impz}. To obtain the result for $C_{\hat{\omega}_k,\hat{\delta}_\ell}(\tau)$, one can use the fact that $\hat{\omega}_k(t) = \frac{d}{dt} \hat{\delta}_k(t)$ to show the relation between the two cross-correlations.
\end{IEEEproof}

{Proposition \ref{prop:ang} leads to a corollary on recovering the frequency response based on its relation to angle response as in \eqref{eq:freq_impz}-\eqref{eq:ang_impz}.}
\begin{corollary}{\emph{(Frequency Response)}}\label{col:freq_ang}
Under (AS\ref{assump1})-(AS\ref{assump3}), the cross-correlation of ambient angle $\hat{\delta}_k$ and $\hat{\delta}_\ell$ is related to the frequency response as
\begin{align}
 T_{u_k,\omega_\ell}(\tau) = -\frac{2\gamma}{\alpha}\frac{d^2}{d\tau^2}C_{\hat{\delta}_k,\hat{\delta}_\ell}(\tau) &= -\frac{2\gamma}{\alpha}\frac{d}{d\tau}C_{\hat{\omega}_k,\hat{\delta}_\ell}(\tau). 
\label{eq:freq_resp2}
\end{align}
\end{corollary}

Proposition \ref{prop:ang} and Corollary \ref{col:freq_ang} nicely extend the ambient frequency data analysis to that for ambient angle data. The key difference is the differentiation needed for achieving the original model coefficients in \eqref{eq:freq_impz}-\eqref{eq:ang_impz}. Similar to \eqref{eq:freq}, there exists a scaling difference between the cross-correlation and dynamic response, which will be discussed in Remark \ref{rmk:scale} soon. 

As mentioned in Remark \ref{rmk:state}, PMUs directly measure the output variables on the bus or branch, instead of the internal angle and speed of generators. Hence, it remains to generalize the cross-correlation equivalence to include actual PMU measurements, namely the bus angle $\theta_n$ and the line power flow $p_{nm}$ from bus $n$ to $m$. To this end, recall that the small-signal analysis framework approximates the output $\bby$ in \eqref{eq:dae} as linear transformation of the state vector $\bbx$. Specifically, the linearized power flow equation in \eqref{eq:swing1} admits that the output bus angle is linearly related to the state as
\begin{align}
	\theta_n(t) =\bba_{n}^\top \bbdelta(t) = \sum_{\ell=1}^N a_{n\ell} \delta_\ell(t),
	\label{eq:lin_bus_ang}
\end{align}
and line flow $p_{nm}(t)$ has similar structures. This linearity is instrumental for extending the analysis to cross-correlating the ambient angle measurement $\hat{\theta}_n(t)$.  This is because the linear relation carries over to the dynamic response and cross-correlation in a similar manner; that is, 
\begin{align}
	T_{u_k,\theta_{n}}(\tau) &= \textstyle \sum_{\ell} a_{n\ell} T_{u_k,\delta_{\ell}}(\tau),\\ 
	{C}_{\hat{\omega}_k,\hat{\theta}_n}(\tau) &= \textstyle\sum_{\ell} a_{n\ell} C_{\hat{\omega}_k,\hat{\delta}_{\ell}}(\tau). 
\end{align}
Note that the ambient $\hat{\omega}_k$ is still not directly measured. As discussed in Remark \ref{rmk:state}, the generator bus frequency, or equivalently the derivative of generator bus angle, can well represent the connected rotor speed. For brevity, we use $\hat{\theta}_k(t)$ as the observed angle at the bus closest to input $u_k$ and establish the following result using the observed ambient angle (frequency).  
\begin{proposition}{\emph{(Bus Angle Response)}} \label{prop:bus_angle}
Under (AS\ref{assump1})-(AS\ref{assump3}), the cross-correlation of ambient bus angle data $\hat{\theta}_k$ and $\hat{\theta}_n$ is related to the bus angle response as
\begin{align}
T_{u_k,\theta_{n}}(\tau) = -\frac{2\gamma}{\alpha}\frac{d}{d\tau} {C}_{\hat{\theta}_k,\hat{\theta}_n}(\tau) = -\frac{2\gamma}{\alpha}{C}_{\hat{\omega}_k,\hat{\theta}_n}(\tau). \label{eq:bus_angle} 
\end{align}
\end{proposition}
Similar to bus angle, the ambient line flow measurements can be used to recover its response as well. 
\begin{proposition}{\emph{(Line Flow Response)}}\label{prop:flow}
Under (AS\ref{assump1})-(AS\ref{assump3}), the cross-correlation of ambient line flow $\hat{p}_{nm}$ and angle $\hat{\theta}_k$ is related to the line flow response as
\begin{align}
T_{u_k,p_{nm}}(\tau) = -\frac{2\gamma}{\alpha}\frac{d}{d\tau} {C}_{\hat{\theta}_k,\hat{p}_{nm}}(\tau) = -\frac{2\gamma}{\alpha}{C}_{\hat{\omega}_k,\hat{p}_{nm}}(\tau). \label{eq:freq_flow}
\end{align}
\end{proposition}

\begin{remark}{\emph{(Scaling Difference)}}\label{rmk:scale}
All the equivalence results between the dynamic responses and cross-correlation outputs have the same scaling coefficient that depends on the input noise level. Hence, it may be difficult to obtain this coefficient from the ambient PMU data itself. Under this scaling difference, the cross-correlation can still be used for determining the time for recovering from the disturbance, or the arresting period \cite{standard2012background}. Moreover, this scaling could also be estimated based on the past event analysis by characterizing the frequency nadir as a function of the disturbance level \cite{peydayesh2017simplified}. 
\end{remark}

\begin{remark}{\emph{(Generalizability)}}\label{rmk:model} 
	Although the three assumptions are necessary for our analytical equivalence results, they can be relaxed to match the practical grid conditions. A key premise for our analysis is that under (AS\ref{assump3})  the  modes are equally and independently excited, such that the cross-correlation output would maintain the same coefficients for all the modes. In practice, the inter-area modes are more evident than local intra-area modes in a wide-area interconnection \cite[Ch.~10]{chow2013power}. As long as the dominant inter-area modes are equally excited, the equivalence results should hold as well. Our numerical studies have demonstrated the cross-correlation outputs can approximately recover the dynamic responses when (AS1)-(AS3) are violated, {\ml including using higher-order generator dynamics and perturbing load demands instead of generator inputs for ambient conditions.}
\end{remark}

\section{The Recovery Algorithm}
\label{sec:algorithm}

\begin{figure}[tb!]
  \centering
  \includegraphics[width=\linewidth]{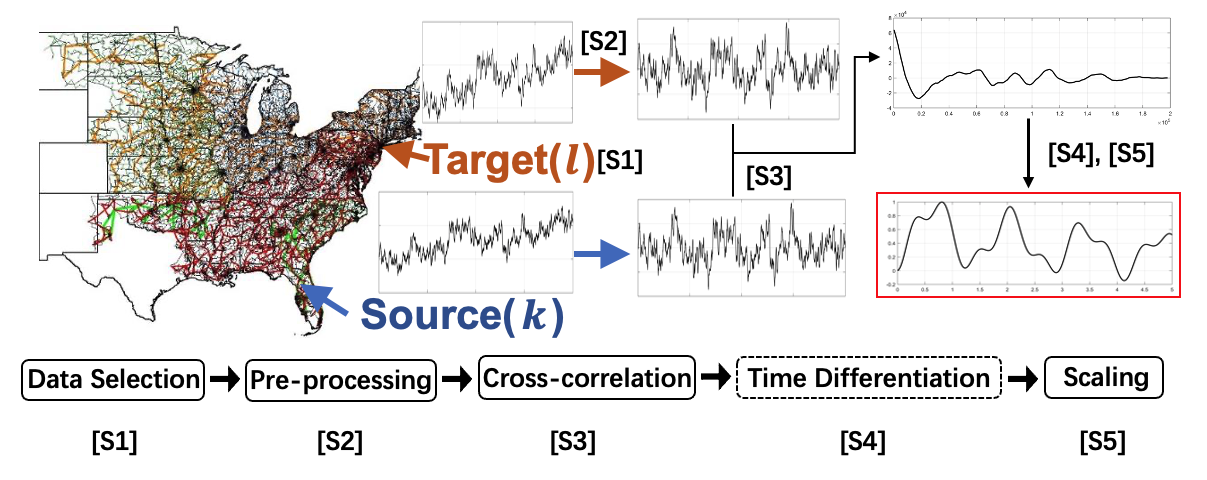}
  \caption{The proposed 5-step algorithm to recover the dynamic responses using ambient synchrophasor data at any two  locations (source and target).}
  \label{fig:alg}
\end{figure}

Based on the analytical equivalence results, we have developed the inference algorithm for recovering the dynamic responses. The implementation is flexible in the types or locations of data.  
Typically, PMUs are installed at critical  substations with large generation or power flow within each control area. As shown by the cross-correlation equivalence results, the PMU data streams from any two locations can be used for implementing the following five-step dynamics recovery algorithm  as illustrated in Fig. \ref{fig:alg}.

\begin{enumerate}
    \item[\textbf{[S1]}] \textbf{(Data Selection)}  For recovering the response $T_{k,\ell}(\tau)$, select the raw data at any source (${x}_k$) and any target (${x}_\ell$) locations, from the closest PMUs in electrical distance as described in Remark \ref{rmk:state} \cite{cotilla2013multi}. For the input generator $k$, this could be frequency/angle data from the substation directly connected to it or a neighboring substation connected through a short transmission line. For the target location, it can be system-level outputs such as bus frequency/angle or line flows. 
      
    \item[\textbf{[S2]}] \textbf{(Pre-processing)} Pre-process the raw data to obtain the proper ambient response signals.  For any angle data, one needs to first find its difference from a reference angle or form the corresponding frequency data using differentiation. The reference angle can be based on the substation connected to the largest generation or the center-of-mass angle \cite{slack_bus} by averaging over all available angle data. Furthermore, we use a bandpass filter to find the detrended signals $\hat{x}_k$ and $\hat{x}_\ell$. 
    As inter-area oscillation modes are of high interest, the passband of the filter is selected accordingly to be {\ml $[0.1,~0.7]~$Hz, which is the typical range of dominant inter-area modes}; see e.g., \cite{wang2015data}. This way, the slowly varying component close to $~0$Hz and fast local modes are filtered out, ensuring to obtain the zero-mean ambient response containing the relevant oscillation modes. 
    
    \item[\textbf{[S3]}] \textbf{(Cross-correlation)}  With the detrended signals at sampling period $T_s$, compute the discrete-time version of the cross-correlation as
          \begin{align*}
          C_{k, \ell}[\tau] = \frac{1}{\mathcal{M}}\sum_{m=1}^{\mathcal{M}}\hat{x}_k[m]\hat{x}_\ell[m-\tau]
          \end{align*}
          where $\mathcal{M} = \lfloor T/T_s\rceil$ is the total number of  samples after rounding.
     
  \item[\textbf{[S4]}] \textbf{(Time Differentiation)} Take the numerical difference of $C_{k, \ell}[\tau]$ depending on the type of dynamic responses of interest. For example, to recover the frequency response from ambient angle data, twice-differentiation may be needed (cf. Corollary \ref{col:freq_ang}).   
  
    \item[\textbf{[S5]}] \textbf{(Scaling)}  If the frequency nadir point is known, one can scale the cross-correlation output to match it. For example, this scaling coefficient could be estimated from past disturbance event analysis as discussed in Remark \ref{rmk:scale}. Otherwise, the recovered responses will be used for evaluating the propagation time. 
\end{enumerate}

Note that the proposed algorithm can directly be applied to infer any dynamical responses from any source to another target using the available ambient measurements at the two locations. It is very computationally efficient. For $\mathcal{M}$ samples, the computation is mainly due to [S3] at $\ccalO(\mathcal{M}^2)$ \cite{hale2006efficient}. This can be further reduced if one is only interested in a shorter duration of the output $C_{k,\ell} (\tau)$. 
In practice, the scaling factor can be recovered by checking the nadir point of a specific disturbance event by historical data or off-line transient stability studies. 


\begin{figure}[!t]
	\centering
	\includegraphics[width=80mm]{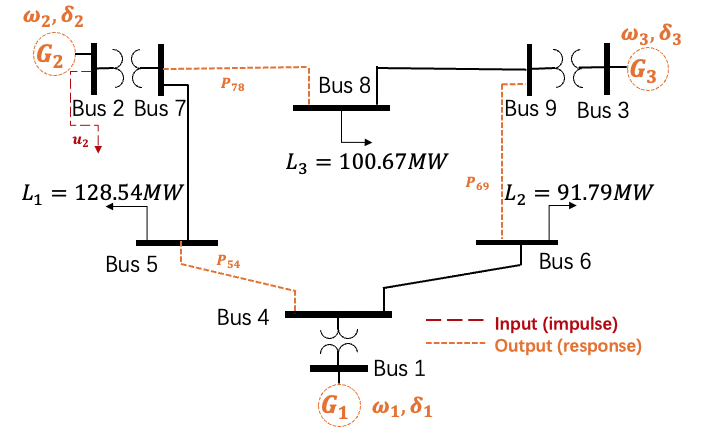}
	\caption{{\ml Diagram of the WSCC 9-bus test case with the impulse input $u_2$ at bus 2. Ambient signals have been generated by perturbing all three loads.}}
	\label{fig:diagram_wscc9}
\end{figure}


\begin{figure*}[!t]
	\centering
	\vspace*{-5mm}
	\includegraphics[width=160mm]{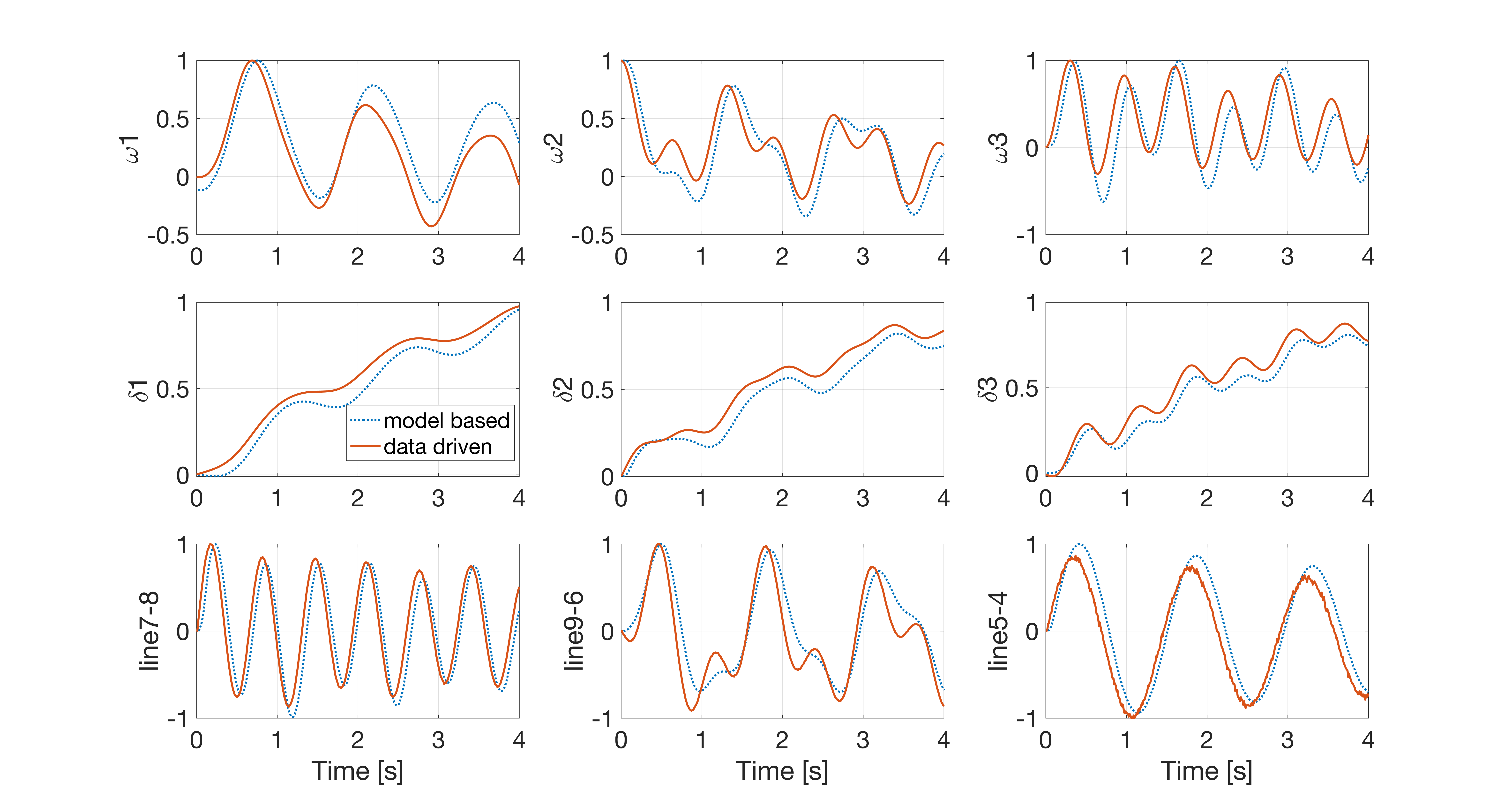}
	\vspace*{-5mm}
	\caption{ {\ml Comparison of model-based and data-driven dynamic responses for the WSCC 9-bus test case under the second-order generator model, uniform damping, and random load perturbations. The disturbance input is set to be $u_2$ at generator bus 2, while the responses are compared for frequency (first row) rotor angle (second row) at all 3 generators, as well as three selected line flows (last row).}}
	\vspace*{-5mm}
	\label{fig:2nd_unif_dyn_rsp_load}
\end{figure*}

\begin{figure*}[!t]
	\centering
	\includegraphics[width=160mm]{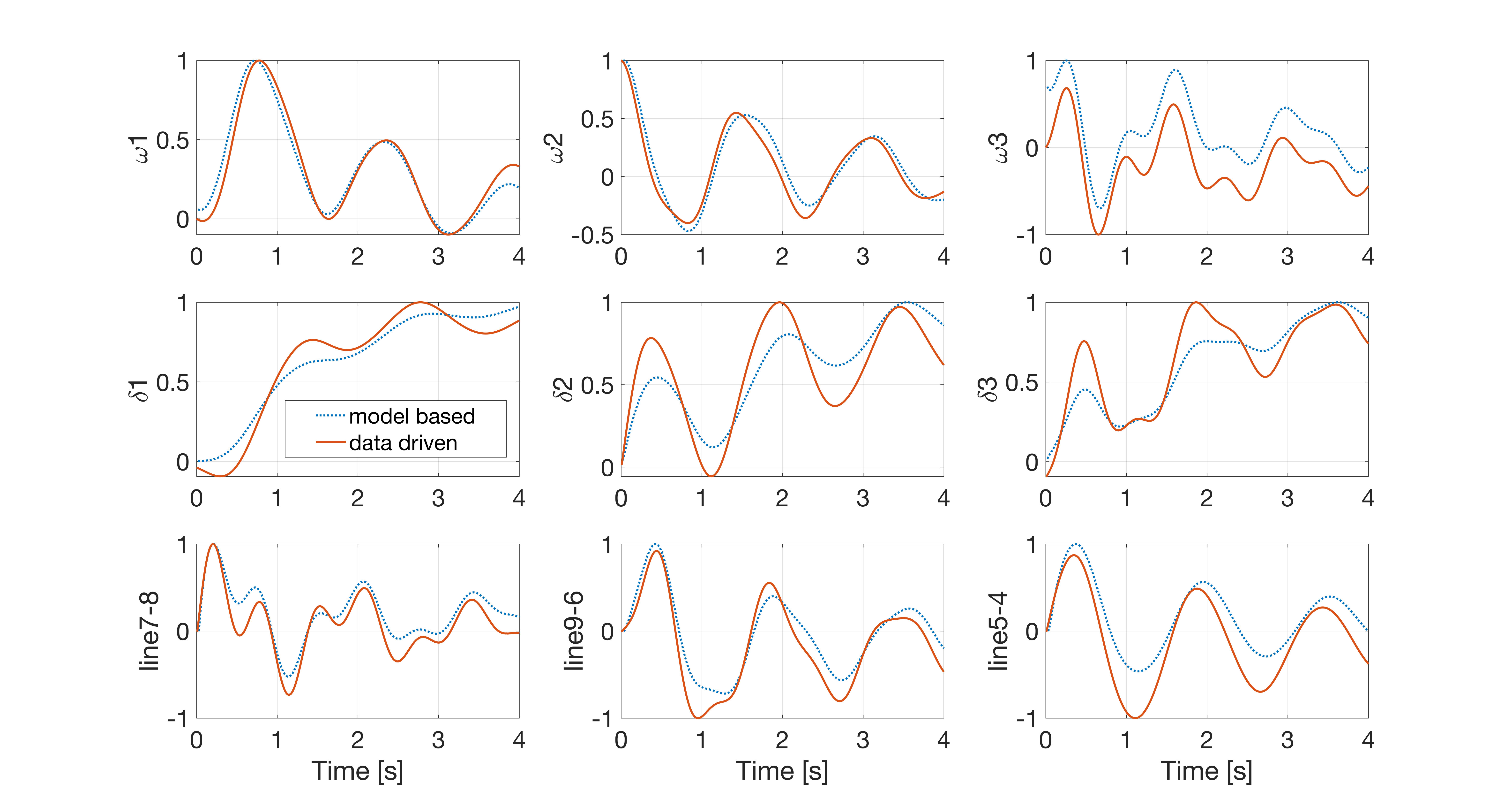}
	\vspace*{-5mm}
	\caption{   {\ml Comparison of model-based and data-driven dynamic responses for the WSCC 9-bus test case under the sixth-order generator model, non-uniform damping, and random load perturbations. The disturbance input is set to be $u_2$ at generator bus 2, while the responses are compared for frequency (first row) rotor angle (second row) at all 3 generators, as well as three selected line flows (last row).}}
	\vspace*{-5mm}
	\label{fig:6th_nonunif_dyn_rsp_load}
\end{figure*}

\section{Numerical Validations}
\label{sec:numerical_results}

This section presents the numerical validation results for the proposed inference algorithms\footnote{The codes and datasets, and results are available at: \newline \indent \url{https://github.com/ShaohuiLiu/dy_resp_pkg_new}}. We use simulated ambient data generated for the WSCC 9-bus system \cite{milano2005open} to validate that the proposed algorithm. Notably, this test case is lossy and includes governor and excitation components that go beyond second-order dynamics, corroborating the generalizability of the proposed method even if assumptions (AS\ref{assump1})-(AS\ref{assump3}) fail to hold. Furthermore, we consider the synthetic ambient synchrophasor data generated from a large 2000-bus system to corroborate the importance of using angle/line flow data over frequency in actual power systems. The 2000-bus case is a realistic  representation of  the Texas grid using actual generator and load models and its dynamic responses have been validated by comparing with actual PMU data; see  details in \cite{idehen2020large}.


\subsection{WSCC 9-Bus System}
\label{subsec:2nd_unif}
The WSCC 9-bus case is widely used for power system dynamic studies \cite{al2000voltage}.  This system has 3 generators, 3 loads and 9 transmission lines, with the one-line diagram shown in Fig.~\ref{fig:diagram_wscc9}. 
Both the impulse responses and ambient data are generated by PSAT \cite{milano2005open} using Matlab.  Matrix $\bbM$ in \eqref{eq:swing2} is given by the test case, while the damping $\bbD$ is set up to demonstrate the need for (AS\ref{assump1}).{\ml To match the realistic grid operations, we set the disturbance location at one generator bus and use load perturbations for ambient data generation.  For the impulse response, we have run time-domain simulation based on the nonlinear DAEs \eqref{eq:dae} with a very short ``impulse''-like input $u_2$ at the generator bus 2. To generate the ambient signals, all the loads have been perturbed with random white noises using the Matlab function \texttt{randn} to mimic (AS\ref{assump3}). Basically, we  (AS3).. }
The sampling rate is set to be very high at 100Hz using the simulation time-step of $dt=0.01s$. {\ml To quantify the recovery accuracy, the metric of normalized mean squared error (MSE) will be used, given by
\begin{align}
 \frac{\| T_{u_k,x_n} - {C}_{k,n}\|_2}{\| T_{u_k,x_n}\|_2} 
\label{eq:est_err}
\end{align}
where $ T_{u_k,x_n} $ ($ {C}_{k,n}$) stands for the model-based (estimated) response normalized by its maximum absolute value.}

We first validated the proposed algorithm under the classical second-order generator model and uniform damping condition. Specifically, the damping coefficient is set to be $\gamma = 0.2$ and thus $\bbD = 0.2 \bbM$.  Note that the power flow Jacobian $\bbK$ is not perfectly symmetric as needed in (AS\ref{assump2}), as the transmission lines are not purely inductive. However, the line resistance-to-reactance (R/X) ratio is very small almost everywhere, and thus $\bbK$ is nearly symmetric. Despite this slight violation of (AS\ref{assump2}), we have observed that the recovered dynamic responses match very well with the model-based impulse responses. Fig.~\ref{fig:2nd_unif_dyn_rsp_load} compares the two for frequency, angle and line flow outputs at different locations. Note that both curves have been normalized by their respective maximum absolute values to eliminate the scale difference.  Clearly, the match between the two is very perfect, except for some small mismatches in the peak values, primarily due to model linearization and asymmetric matrix $\bbK$.

We have further validated the proposed algorithm by retaining the original case settings of sixth-order generator model that includes controllers like governor, exciter, and power system stabilizer. We also changed the damping to be non-uniform ($\gamma\in [0.1,0.3]$). 
All these settings reflect the realistic power system dynamics with nonlinearity and ambient conditions. 
Fig.~\ref{fig:6th_nonunif_dyn_rsp_load} plots the updated comparisons for this case, which further confirm the effectiveness of our proposed data-driven approaches in recovering the dynamic responses. Compared to Fig.~\ref{fig:2nd_unif_dyn_rsp_load}, the dynamic responses exhibit varying modal components and better damping effects, especially for the frequency responses. Nonetheless, the proposed data-driven approach can still well capture the transients therein, despite that this test has significantly deviated from our assumptions (AS\ref{assump1})-(AS\ref{assump2}). 


Table \ref{table:wscc9-accuracy} lists the normalized MSE in \eqref{eq:est_err} for recovering dynamic responses, by averaging over all system locations. Interestingly, the recovery performance is largely the same between the two different cases of generator models and damping conditions, except for the line flow responses. Therefore, relaxing our analytical assumptions to more realistic grid conditions has led to minimal effect on the recovery performance.  
Overall speaking, the proposed framework can well recover different types of system responses, corroborating its effectiveness and generalizability  based on simulated tests.

\begin{table}[!t]
\centering
\caption{{\ml Normalized MSE of recovering different dynamic responses for the WSCC 9-bus test case.} }
\begin{tabular}{lll}
\hline
& 2nd-unif & 6th-nonunif \\ \hline
Frequency & 0.25    & 0.26   \\
Angle      & 0.12 & 0.10 \\
Flow       & 0.20    & 0.37  \\  [1ex] 
\hline \\
\end{tabular}
\label{table:wscc9-accuracy}
\end{table}


 \subsection{2000-Bus Synthetic Texas System}


Due to the limited access to actual synchrophasor data, we have utilized the synchrophasor data generated for the 2000-bus synthetic Texas grid \cite{birchfield2016grid} for large system validations. 
{\ml This system has 1500 substations, 500 generators, approximately 50,000 MW of peak load, and 3206 transmission lines. The dynamic component models including load models follow from an actual Texas system model, and the dynamic responses have been validated with the actual ones. The ambient signals are generated as follows \cite{idehen2020large}: i) Periodic variations at 5s- and 7s-intervals have been set up for loads and generators, respectively; ii) A simulation time step of one quarter cycles, with power flow result storage every 8 time steps, is used by the solver specification, close to PMU data rates of 30 samples per second. This test will use a total duration of $10$min data with a sampling rate of $30$Hz. The simulated ambient data have been further processed to produce synthetic synchrophasor data, by adding {0.002\% random measurement noises to all data streams}. In addition, the frequency data are further filtered based on the actual PMU processing method described in \cite{idehen2020large}, which can mimic the statistical property of actual data. The filtering process has made the synthetic ambient frequency data to be unreliable for recovering dynamic responses, as detailed soon. }

\begin{figure}[!t]
  \centering
  \vspace*{-5mm}
  \includegraphics[width=.6\linewidth]{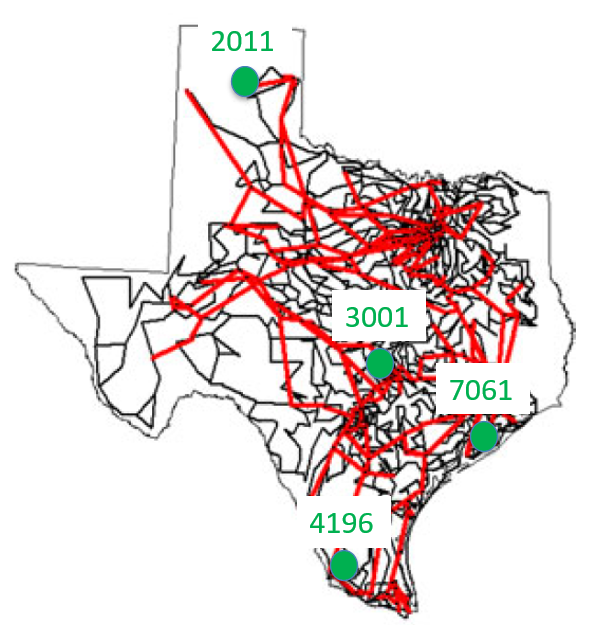}
 	\vspace*{-5mm}
  \caption{The network topology model for the 2000-bus synthetic Texas grid \cite{birchfield2016grid}. There are a total of 99 PMUs placed in the system, all at 345kV buses with red lines indicating these high-voltage transmission lines as well. The remaining gray lines correspond to 115kV lines. }
  \label{fig:texas2000}
\end{figure}

Fig.~\ref{fig:texas2000} illustrates this synthetic grid system, which is a validated replica of the ERCOT system. The distance between the north region Bus 2011 and south region Bus 4196 is around 670 miles, while the distance from north to coast region Bus 7061 is about 450 miles. The frequency responses are very similar within the system due to the system size and frequency control designs \cite[Ch.~10]{ercot_control2016}. To compare the responses, we have picked Bus 2011 in the north region as the location of the input source, and other three buses (Buses 3001, 4196, and 7061) as the output target locations. 

\begin{figure}[t!]
    {
        \centering
        \includegraphics[width=90mm]{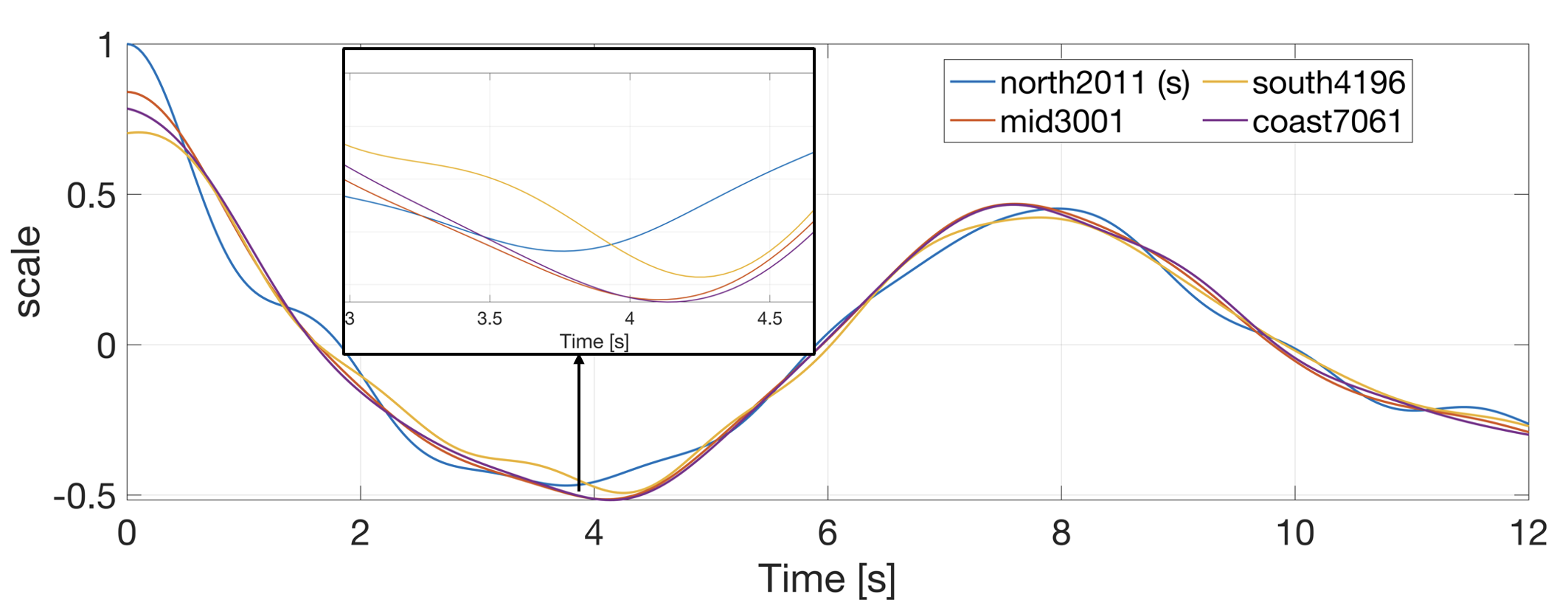}
        \centerline{(a) Simulated frequency data}
        \hspace*{-5mm}
        \includegraphics[width=100mm]{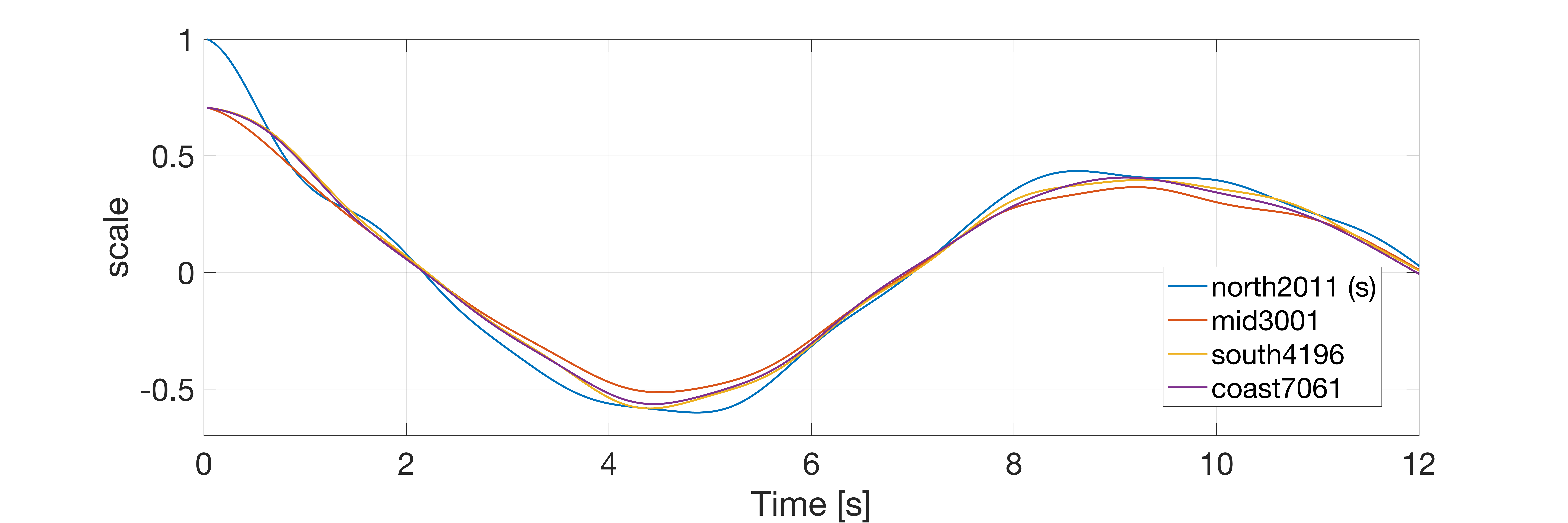}
        \centerline{(b) Synthetic angle data}
        \hspace*{-5mm}
        \includegraphics[width=100mm]{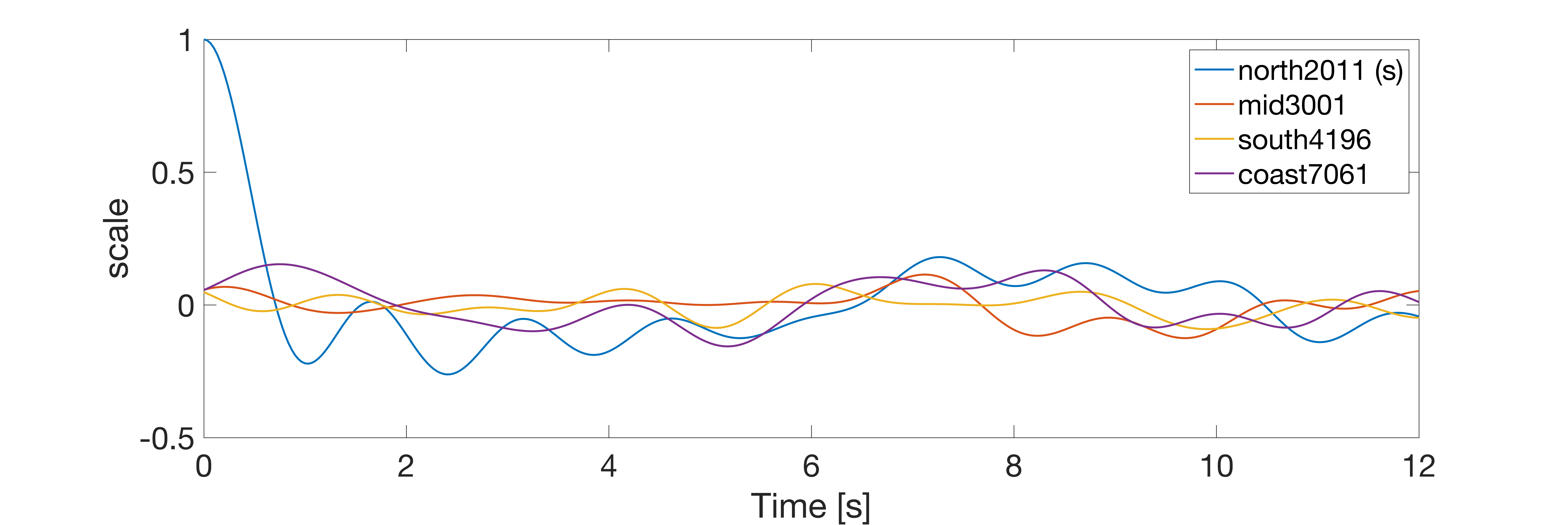}
        \centerline{(c) Synthetic frequency data}
        \caption{{\ml Comparison between recovered frequency responses from (a) simulated frequency data; (b) synthetic angle data; and (c) synthetic frequency data at four selected locations, with input disturbance at Bus 2011 in the north region.}}
        \label{fig:freq_resp}}
\end{figure}

We first compare the recovered frequency responses obtained by both ambient frequency data and angle data, as plotted in Fig.~\ref{fig:freq_resp} . 
To process the ambient data, we have set the filter pass bands to be {\ml $[0.1,~0.7]$Hz} as in \textbf{[S2]}. Specifically for angle data, we first compute a reference angle by taking the average over all recorded angle data within the system, as discussed in \textbf{[S1]}. After obtaining the reference angle, we subtract it from the ambient angle data  before using the bandpass filter. Moreover, a final step to process angle data is to take the time differentiation of the cross-correlation output  [cf. \eqref{eq:bus_angle}] in order to recover the angle responses, or a twice time differentiation for the frequency responses.  
{\ml The simulated frequency data have been used as the benchmark for evaluating both synthetic angle and frequency data.  Using the simulated frequency data,} the proposed cross-correlation outputs show very similar frequency responses at all locations, except for some  minor time lags among the first nadir points, as shown by Fig.~\ref{fig:freq_resp}(a). The time of frequency nadir points as estimated by our proposed algorithm is listed in Table \ref{table:freq_lag}. A closer look at the time lags confirms with the nominal speed for electromechanical wave propagation, which is around 200-1,000~mi/sec for typical systems \cite{alharbi2020simulation}. {\ml 
The synthetic angle data produce very similar frequency responses in Fig.~\ref{fig:freq_resp}(a), corroborating the effectiveness of the proposed general framework. However,  due to PMUs' signal processing step in filtering frequency data, the synthetic frequency data have led to highly inaccurate frequency responses which clearly lack in synchronization, as shown in the Fig.~\ref{fig:freq_resp}(c). This comparison speaks for the importance of the proposed extension over \cite{huynh2018data} which used frequency data only. }

\begin{table}[!t]
\centering
\caption{The time of nadir points and their lags at the four locations in the 2000-bus Texas system along with the propagation speed, as estimated by the proposed cross-correlation approach. }
\begin{tabular}{lllll}
\hline
Bus Index   & 2011 & 3001 & 7061 & 4196 \\ \hline
Distance/mi & 0    & 370  & 535  & 670  \\
Time/s      & 3.73 & 4.06 & 4.16 & 4.27 \\
Lag/s       & 0    & 0.33 & 0.43 & 0.54 \\  [1ex] 
\hline \\
\end{tabular}
\label{table:freq_lag}
\end{table}



\begin{figure}[!t]
  \centering
  \includegraphics[width=90mm]{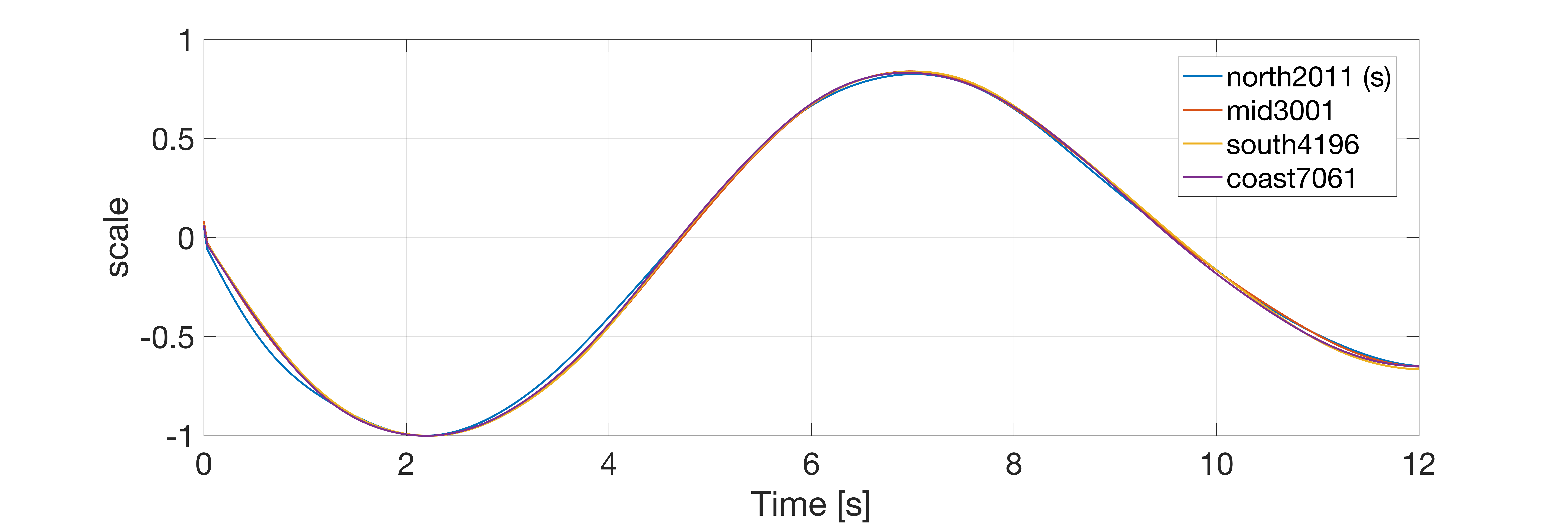} 
  \vspace*{-5mm}
  \caption{{\ml Recovered angle responses from ambient angle data at four selected locations with input disturbance from Bus 2011 in the north region.}}
  \vspace*{-5mm}
  \label{fig:ang_resp}
\end{figure}

We have further evaluated the recovery of angle responses using the synthetic angle  data, as plotted in Fig.~\ref{fig:ang_resp}.  Similar to the frequency responses, the underlying oscillation modes in  all the angle responses lines are very similar. 
To sum up, the proposed data-driven algorithm has been shown very accurate in recovering the grid dynamic responses using synthetic ambient synchrophasor data that have been realistically generated for the 2000-Bus Texas system.  Its practical value has been demonstrated by the reliable recovery performance thanks to the higher accuracy of  ambient angle data over frequency data.




\section{Conclusions} 
\label{sec:con}

This paper develops a general data-driven framework for recovering small-signal dynamic responses from various types of ambient synchrophasor data. We have proposed a cross-correlation based technique to process the ambient data from any two locations of interest. We have analyzed the second-order dynamic models in the small-signal regime, and theoretically establish the equivalence between the proposed cross-correlation result and the inertia-based dynamic responses under some mild assumptions that hold for large-scale power systems. This equivalence allows to develop a general framework to process ambient frequency, angle, line flow data, that is flexible in the locations and types of PMU data. Numerical tests on the WSCC 9-bus case have demonstrated that the effectiveness of the proposed cross-correlation technique, even in the presence of higher-order dynamics and violated assumptions as in realistic power systems. Additional tests using synthetic synchrophasor data generated for a realistic 2000-bus Texas system also strongly support our algorithm for large system implementation. Thus, our proposed data-driven technique provides a general framework for processing ambient synchrophasor data to recover dynamic responses without requiring the grid modeling information.

Exciting new  directions open up, such as the integration with model estimation methods to recover the system parameters, as well as the consideration of high-dimensional data analysis tools to utilize system-wide measurements. Furthermore, we are actively exploring example applications for the proposed framework in terms of evaluating the effectiveness of advanced control designs and data-driven modeling in weak grids with high penetration of inverter-based resources.


%
\bibliographystyle{IEEEtran}

\itemsep2pt
\bibliography{ref.bib}

\end{document}